\documentclass[
reprint, 
noeprint,
superscriptaddress,
amsmath,amssymb,
aps,
prl,
twocolumn,
]{revtex4-1}
 
\usepackage{graphicx}% Include figure files
\usepackage{dcolumn}% Align table columns on decimal point
\usepackage{bm}% bold math
\usepackage{hyperref}% add hypertext capabilities
\usepackage{comment, multirow, subfigure, longtable}
\usepackage{color}
\usepackage{url}
\usepackage[left]{lineno}
\usepackage{makecell}

\begin{document}
%%%%%%%%%%%%%%%%%%%%%% TABLE 1 %%%%%%%%%%%%%%%%%%%%%%

\newcommand{\tabnumberofevents}{
	\begin{table}[h]
		\centering
		\begin{tabular}{lcccccc}
			\hline
			\hline
			& 2008 & 2009 & 2010 & 2011 & 2012 & Total\\
			\hline
			%p.o.t. ($10^{19}$) & 1.74 & 3.53 & 4.09 & 4.75 & 3.86 & 18\\
			p.o.t. ($10^{19}$) & 1.74 & 3.53 & 4.09 & 4.75 & 3.86 & 17.97\\
			$0\mu$ events & 150 & 255 & 278 & 291 & 223 & 1197 \\
			$1\mu$ events ($p_\mu<15$~GeV/c) & 543 & 1024 & 1001 & 1031 & 807 & 4406 \\
			Total events & 693 & 1279 & 1279 & 1322 & 1030 & 5603 \\
			%$\nu_\tau$ candidates ** & - & 1 & - & 1 & 3 & 10 \\
			\hline
			\hline
		\end{tabular}
		\caption{Number of events used in this analysis, grouped into $0\mu$ and $1\mu$.
		\label{tab:Nevents}}
	\end{table}
}

%%%%%%%%%%%%%%%%%%%%%% TABLE 2 %%%%%%%%%%%%%%%%%%%%%%

\newcommand{\tabselectioncuts}{
  \begin{table}[h!!]
	 \centering
	 \begin{tabular}{lcccc}
		\hline
		\hline
		\textbf{Variable} & \textbf{$\tau\to 1h$} &
		\textbf{$\tau\to 3h$} & \textbf{$\tau\to \mu$} &
		\textbf{$\tau\to e$}\\
		\hline
		\textbf{$z_{dec}$} (mm) & \text{$<2.6$} &
		\text{$<2.6$} & \text{$<2.6$} &
		\text{$<2.6$}\\
		\textbf{$\theta_{kink}$} (rad) & \text{$>0.02$} &
		\text{$>0.02$} & \text{$>0.02$} &
		\text{$>0.02$}\\
		\textbf{$p_{2ry}$} (GeV/c) &  \text{$>1$} & \text{$>1$} & \text{[1,\,15]} &
		\text{$>1$}\\
		\textbf{$p^{T}_{2ry}$} (GeV/c) & \text{$>0.15$} &
		\text{-} & \text{$>0.1$} &
		\text{$>0.1$}\\
		% \textbf{$p^{T}_{miss}\ (GeV/c)$}  & \text{/} & \text{/} & \text{/} & \text{/}\\
		% \textbf{$\phi_{lH}\ (rad)$} & \text{/} & \text{/} & \text{/} &  \text{/}\\
		%  \textbf{$m, m_{min}\ (GeV/c^{2})$} & \multicolumn{2}{c|}{\text{/}} &
		%  \text{[0.5,\,2]} & \text{/} & \multicolumn{2}{c|}{/} & \multicolumn{2}{c}{\text{/}}\\
        charge$_{\textnormal{2ry}}$ & \text{-} & \text{-} &
		\footnotesize 
        \makecell{ negative or\\unknown} & \text{-} \\
		\hline
		\hline
		%\multicolumn{9}{c}{Cuts marked with $\star$ are not applied for Quasi-Elastic event}\\                 
		%\multicolumn{9}{c}{* $p^{T}_{2ry}$ cut is 0.3 in the presence of $\gamma$ particles associated to the decay vertex}
	\end{tabular}
	\caption{Selection cuts.\label{tab:KinSel}} 
  \end{table}
}

%%%%%%%%%%%%%%%%%%%%%% TABLE 3 %%%%%%%%%%%%%%%%%%%%%%

\newcommand{\tabnexpectedevents}{
  \begin{table*}[t]
	\centering%\footnotesize
	\begin{tabular}{c|cccc|c|c}
		\hline
		\hline
		\textbf{Channel} & \multicolumn{4}{c|}{\textbf{Expected Background}} & \textbf{$\nu_\tau$ Exp.}& \textbf{Observed}\\
		& Charm & Had. re-interaction & Large $\mu$-scat. & Total & \textbf{}\\
		\hline
		$\tau\to 1h$ & $0.15 \pm 0.03$ & $1.28\pm 0.38$ & $-$ & $1.43\pm 0.39$ & $2.96\pm 0.59$ & 6\\
		$\tau\to 3h$ & $0.44\pm 0.09$ & $0.09\pm 0.03$ & $-$ & $0.52\pm 0.09$ & $1.83\pm 0.37$ & 3 \\
		$\tau\to\mu$ & $0.008\pm 0.002$ & $-$ & $0.016\pm 0.008$ & $0.024\pm 0.008$ & $1.15\pm 0.23$ & 1\\
		$\tau\to e$ & $0.035\pm 0.007$ & $-$ & $-$ & $0.035\pm 0.007$ & $0.84\pm 0.17$ & 0\\
		\hline
		\textbf{Total} & $0.63\pm 0.10$ & $1.37\pm 0.38$ & $0.016\pm 0.008$ & $2.0\pm 0.4$ & $6.8\pm 0.75$ & 10\\
		\hline
		\hline
	\end{tabular}
	\caption{The expected number of signal and background events for the analysed data sample, evaluated assuming ${\Delta m^{2}_{23} = 2.5\cdot 10^{-3}\ \textnormal{eV}^{2}}$, ${\sin^{2}2\theta_{23} = 1}$ and the default implementation for the $\nu_\tau$ cross-section of GENIE~v2.6.
	\label{tab:NExp_MinBias}}
  \end{table*}
}

%%%%%%%%%%%%%%%%%%%%%% TABLE 4 %%%%%%%%%%%%%%%%%%%%%%

\begin{comment}
\newcommand{\tabkinematicalvars}{
  \begin{table*}[t]
	\centering
	\begin{tabular}{c|ccccc}
		\hline
		\hline
		Event ID & 11143018505 & 11172035775 & 9190097972 & 10123059807 & 11213015702 \\
		\hline
		% Event Class. &  $0\mu$ &  $0\mu$ &  $0\mu$ &  $0\mu$ &  $0\mu$\\
		Topology & 1-prong & 1-prong & 1-prong & 3-prong & 3-prong \\
		$z_{dec}$ ($\mu$m) & 430 & 652	& {10} & -648 & 407 \\
		$p^{T}_{miss}$ (GeV/c) & 0.88 & {1.29} & 0.46 & 0.60 & $>0.50$\\
		$\phi_{lH}$ (degrees)&152 & 140 & 143 & {82} & {47} \\ 
		$p^T_{2ry}$ (GeV/c)& {0.24} & {0.25} & {0.33} & / & /\\
		$p_{2ry}$ (GeV/c)& 2.7 & 2.6 & 2.2 & 6.7 & $>6.3$\\
		$\theta_{kink}$ (mrad)& 90 &  98 & 146 & 231 & 83 \\
		$m$ (GeV/c$^2$)& / & / & / & 1.2 & {$>0.94$} \\
		$\gamma$ at decay vtx & 1 & 0 & 0 & 0 & 2\\            \hline
		\hline
	\end{tabular}
	\caption{Kinematical variables for the newly identified $\nu_\tau$ candidates. The other five $\nu_\tau$ candidates were described in ~\cite{OPERA:1tau,OPERA:2tau,OPERA:3tau,OPERA:4tau,OPERA:5tau}}
	\label{tab:MarginalEv}
  \end{table*}
}
\end{comment}

\newcommand{\tabkinematicalvars}{
  \begin{table*}[t]
	\centering
	\begin{tabular}{l|c c c c c c c c c c}
		\hline
		\hline
		%Event ID & 9234119599 & 11113019758 & 12123032048 & 12254000036 & 12227007334 & 11143018505 & 11172035775 & 9190097972 & 10123059807 & 11213015702 \\
        BRICK ID & 72693 & 29570 & 23543 & 92217 & 130577 & 77152 & 27972 & 26670 & 136759 & 4838\\
		\hline
		% Event Class. &  $0\mu$ &  $0\mu$ &  $1\mu$ &  $0\mu$ &  $0\mu$ &  $0\mu$ &  $0\mu$ &  $0\mu$ &  $0\mu$ &  $0\mu$\\
		%\textit{Topology} & 1-prong & 3-prong & 1-prong & 3-prong & 3-prong & 1-prong & 1-prong & 1-prong & 3-prong & 3-prong \\
        \textit{Channel} & $\tau\to1h$ & $\tau\to3h$ & $\tau\to\mu$ & $\tau\to1h$ & $\tau\to1h$ & $\tau\to1h$ & $\tau\to1h$ & $\tau\to1h$ & $\tau\to3h$ & $\tau\to3h$\\ 
		$z_{dec}$ ($\mu$m) & 435 & 1446	& 151 & 406 & 630 & 430 & 652	& {303} & -648 & 407 \\
		$p^{T}_{miss}$ (GeV/c) & 0.52 & {0.31} & / & 0.55 & 0.30 & 0.88 & {1.29} & 0.46 & 0.60 & $>0.50$\\
		$\phi_{lH}$ (degrees) & 173 & 168 & / & 166 & 151 &152 & 140 & 143 & {82} & {47} \\ 
		$p^T_{2ry}$ (GeV/c) & 0.47 & / & 0.69 & 0.82 & 1.00 & {0.24} & {0.25} & {0.33} & / & /\\
		$p_{2ry}$ (GeV/c)  & 12 & 8.4 & 2.8 & 6.0 & 11 & 2.7 & 2.6 & 2.2 & 6.7 & $>6.3$\\
		$\theta_{kink}$ (mrad)& 41 &  87 & 245 & 137 & 90  & 90 &  98 & 146 & 231 & 83 \\
		$m$ (GeV/c$^2$) & / & 0.80 & / & 1.2 & {$>0.94$} & / & / & / & 1.2 & {$>0.94$} \\
		$\gamma$ at decay vtx & 2 & 0 & 0 & 0 & 0 & 1 & 0 & 0 & 0 & 2\\ 
        charge$_{\textnormal{2ry}}$ & / & / & -1 & / & / & / & / & / & / & /\\ 
        BDT Response & 0.32 &  -0.05 & 0.37 & 0.12 & 0.35 & 0.18 & -0.25 & -0.10 & -0.04 & -0.03\\            
        \hline
		\hline
	\end{tabular}
	\caption{Kinematical variables and BDT response for all $\nu_\tau$ candidates.}
	\label{tab:MarginalEv}
  \end{table*}
}

\newcommand{\authorlist}{
\author{N. Agafonova}
\affiliation{INR - Institute for Nuclear Research of the Russian Academy of Sciences, RUS-117312 Moscow, Russia}
\author{A. Alexandrov}
\affiliation{INFN Sezione di Napoli, I-80126 Napoli, Italy}
\author{A. Anokhina}
\affiliation{SINP MSU - Skobeltsyn Institute of Nuclear Physics, Lomonosov Moscow State University, RUS-119991 Moscow, Russia}
\author{S. Aoki}
\affiliation{Kobe University, J-657-8501 Kobe, Japan}
\author{A. Ariga}
\affiliation{Albert Einstein Center for Fundamental Physics, Laboratory for High Energy Physics (LHEP), University of Bern, CH-3012 Bern, Switzerland}
\author{T. Ariga}
\affiliation{Albert Einstein Center for Fundamental Physics, Laboratory for High Energy Physics (LHEP), University of Bern, CH-3012 Bern, Switzerland}
\affiliation{Faculty of Arts and Science, Kyushu University, J-819-0395 Fukuoka, Japan}
\author{A. Bertolin}
\affiliation{INFN Sezione di Padova, I-35131 Padova, Italy}
\author{C. Bozza}
\affiliation{Dipartimento di Fisica dell'Universit\`a di Salerno and ``Gruppo Collegato''  INFN, I-84084 Fisciano (Salerno), Italy}
\author{R. Brugnera}
\affiliation{INFN Sezione di Padova, I-35131 Padova, Italy}
\affiliation{Dipartimento di Fisica e Astronomia dell'Universit\`a di Padova, I-35131 Padova, Italy}
\author{A. Buonaura}
\altaffiliation{Now at Physik-Institut, Universit{\"a}t Z{\"u}rich, Z{\"u}rich, Switzerland}
\affiliation{INFN Sezione di Napoli, I-80126 Napoli, Italy}
\affiliation{Dipartimento di Fisica dell'Universit\`a Federico II di Napoli, I-80126 Napoli, Italy}
% ...
}
%\preprint{APS/123-QED}
\title{Final results of the OPERA experiment on $\nu_\tau$ appearance in the CNGS beam}
\authorlist
\collaboration{OPERA Collaboration}

\begin{abstract}

The OPERA experiment was designed to study $\nu_\mu\to\nu_\tau$ oscillations in appearance mode in the CNGS neutrino beam. 
In this letter we report the final analysis of the full data sample collected between 2008 and 2012, corresponding to $17.97\cdot 10^{19}$ protons on target. Selection criteria looser than in previous analyses have produced ten $\nu_\tau$ candidate events, thus reducing the statistical uncertainty in the measurement of the oscillation parameters and of $\nu_\tau$ properties. A multivariate approach for event identification has been applied to the candidate events and the discovery of $\nu_\tau$ appearance is confirmed with an improved significance level of 6.1~$\sigma$. $\Delta m^2_{23}$ has been measured, in appearance mode, with an accuracy of 20$\%$. The measurement of $\nu_\tau$~CC cross-section, for the first time with a negligible contamination from $\bar{\nu}_\tau$, and the first direct observation of the $\nu_\tau$ lepton number are also reported.
\end{abstract}

\pacs{14.60.Pq}% PACS, the Physics and Astronomy
                             % Classification Scheme.
%\keywords{Suggested keywords}%Use showkeys class option if keyword
                              %display desired
\maketitle

\section{Introduction\label{sec:intro}}

The OPERA experiment was designed to conclusively prove $\nu_\mu \to \nu_\tau$ oscillations in appearance mode. 
The challenge of the experiment to detect the short-lived $\tau$ lepton (c$\tau$ = 87~$\mu$m), produced in the charged-current~(CC) $\nu_{\tau}$ interactions, was accomplished with the nuclear emulsion technique, that provides sub-micrometric spatial resolution.
\\
The detector~\cite{OPERA_4} was located in the underground Gran Sasso Laboratory (LNGS), 730~km away from the neutrino source and exposed to the CNGS muon neutrino beam~\cite{CNGSbeam:1998, Baldy:1999dc}. 
The average neutrino energy was about 17~GeV, the $\bar{\nu}_{\mu}$ fraction was 2.1$\%$ in terms of expected CC interactions, the sum of $\nu_{e}$ and $\bar{\nu}_{e}$ was below 1$\%$, while the prompt $\nu_{\tau}$ contamination was negligible $\mathcal{O}(10^{-7}$).

The detector was an hybrid apparatus made of an emulsion/lead target with a total mass of about 1.25 kt, complemented by electronic detectors. The general structure consisted of two identical Super Modules (SM). 
Each SM was made of a target section, composed of 31 target walls, and a muon spectrometer. 
Each target wall was an assembly of horizontal trays loaded with target units, hereafter called \textit{bricks}. 
Each brick consisted of 57 emulsion films, 300~$\mu$m thick, interleaved with 56 lead plates, 1~mm thick, with a $(12.7\times10.2)$~cm$^{2}$ cross-section, a thickness of 7.5~cm corresponding to about 10~radiation lengths and a mass of 8.3~kg. 
Downstream of each target wall, two orthogonal planes of scintillator strips (Target Tracker detector) recorded the position and the energy deposition of charged particles~\cite{Adam:2007ex}. Finally, a magnetic spectrometer instrumented with Resistive Plate Chambers and high-resolution Drift Tubes was used to identify muons and measure their charge and momentum~\cite{OPERA_4}. 
Neutrino interactions and decay topologies were detected in the bricks with micrometric accuracy. 
A pair of emulsion films was attached to the downstream face of each brick, acting as an interface between the brick and the electronic detectors~\cite{Anokhina:2008aa}. Their measurements allowed confirming the presence of tracks recorded in the electronic detectors before unpacking and developing the entire brick. A detailed description of the OPERA detector can be found in~\cite{OPERA_4}.

\paragraph*{Event selection and analysis}\label{sec:evsel}

The appearance of the $\tau$ lepton is identified by the detection of its characteristic decay topologies, either in 1-prong (electron, muon or hadron) or in 3 hadron prongs. Kinematical selection criteria are applied to reduce the background coming from the processes that mimic the $\tau$ decay topologies, which are: i) the decays of charmed particles produced in $\nu_{\mu}$~CC interactions; ii) re-interactions of hadrons from $\nu_{\mu}$ events in lead; iii) the large-angle scattering (LAS) of muons produced in $\nu_{\mu}$~CC interactions. Processes i) and $\nu_{\mu}$~CC in ii) represent a background source when the $\mu^-$ at the primary vertex is not identified.

A sample corresponding to $17.97\times10^{19}$ protons on target (p.o.t.) has been collected from 2008 to 2012 and resulted in 19505 neutrino interactions in the target fiducial volume.

In 2015, five $\nu_\tau$ candidates were observed following a selection performed by cuts on specific kinematical and topological parameters. The discovery of $\nu_\mu\to \nu_\tau$ oscillations was assessed with a significance of 5.1~$\sigma$~\cite{OPERA:5tau}.

This paper reports an improved analysis of the full data sample, which is $3.6\%$ higher with respect to~\cite{OPERA:5tau}. The number of fully analysed events is shown in Tab.~\ref{tab:Nevents} for each year of data taking. Events are classified as 1$\mu$ if a track is tagged as a muon by the electronic detectors~\cite{Agafonova:2011zz}, as 0$\mu$ otherwise. 
The analysis described below is performed on all 0$\mu$ events and on 1$\mu$ events with a muon momentum below 15~GeV/c.

\tabnumberofevents
The new analysis is based on a multivariate approach for event identification, fully exploiting the expected features of $\nu_\tau$ events, rather than on the sheer selection of candidate events by independent cuts on topological or kinematical parameters as in previous analyses. It is performed on candidate events preselected with looser cuts than those applied in the previous cut-based approach~\cite{OPERA:1tau,OPERA:2tau,OPERA:3tau,OPERA:4tau,OPERA:5tau}. Looser cuts increase the number of $\nu_\tau$ candidates, thus leading to a measurement of the oscillation parameters and of the $\nu_\tau$ cross-section with a reduced statistical uncertainty. Given the higher discrimination power of the multivariate analysis that fully exploits the features of each event the significance of the $\nu_\tau$ appearance is increased.

\section{Analysis strategy\label{sec:strategy}} 
The first stage of the analysis is to select events showing a decay topology. These events are categorised into 4 channels (\textbf{$\tau\to 1h$}, \textbf{$\tau\to 3h$}, \textbf{$\tau\to \mu$}, \textbf{$\tau\to e$}) according to the identification of daughter particles. Then, kinematical cuts are applied to refine the selection and to reject background events in each channel. Finally, a multivariate approach is exploited to separate signal from background and to evaluate a single-variable discriminant for the hypothesis test and parameters estimation in the statistical analysis for the extraction of results, as described in the next section.
\\
The rectangular cuts on the topological and kinematical variables, shown in Table~\ref{tab:KinSel}, are looser than those used in previous papers~\cite{OPERA:1tau, OPERA:2tau, OPERA:3tau, OPERA:4tau, OPERA:5tau}. 
The Monte Carlo simulation has been validated comparing its results with the measured $\nu_\mu$~CC interactions when producing hadron re-interactions~\cite{Ishida:2014qga}, charmed hadron decays~\cite{Agafonova:2014khd} and LAS muons~\cite{Longhin:2015dsa}. The momentum of hadrons has been estimated by the multiple Coulomb scattering method~\cite{MCS:2011}, while the muon momentum is measured by the magnetic spectrometer with a resolution of about $20\%$~\cite{OPERA_4}.

Decay topologies are identified by the following requirements: the average 3D angle between the parent and its daughters ${(\theta_{kink})}$ has to be larger than 0.02~rad and the distance between the decay vertex and the downstream face of the lead plate containing the primary vertex $(z_{dec})$ has to be shorter than 2600~$\mu$m. To define the decay vertex position with sufficient precision, the total momentum of the visible tracks coming out of the secondary vertex ($p_{2ry}$) has to be at least 1~GeV/c, with an upper limit of 15~GeV/c only for the $\tau\to\mu$ channel, in order to reduce the charmed hadrons background. 
Moreover, for 1-prong decays, the cut on the daughter transverse momentum with respect to the parent direction ($p^{T}_{2ry}$) was tuned to reduce the hadron re-interaction and LAS backgrounds. Lastly, for the $\tau\to\mu$ channel, only events where the muon daughter has negative or unknown charge~\cite{internal_note:2009} are considered. 

\tabselectioncuts

The tracks related to events passing the selection criteria of Table~\ref{tab:KinSel} have been measured within an angular acceptance up to $\tan\theta<1$ ($\theta$ being the angle of the track with the z axis) for kinematical measurements. In addition, a specific search for large angle tracks, up to $\tan\theta<3$, has been performed, in order to reject events with nuclear fragments emitted at the secondary vertex, a signature of the hadronic re-interaction background.

After candidate selection, a multivariate analysis is applied, based on a Boosted Decision Tree (BDT) algorithm implemented in TMVA~\cite{root_tmva}. For each channel, the BDT was trained with Monte Carlo events selected according to the topology and the kinematical cuts of Table~\ref{tab:KinSel}. As input for the BDT analysis, additional kinematical variables have been used. As described in~\cite{OPERA:2tau}, they are: the missing transverse momentum with respect to the incoming neutrino direction (${p^{T}_{miss}}$), the transverse opening angle between the $\tau$ candidate and the hadronic system (${\phi_{lH}}$) and the invariant mass of the parent particle ($m$). In addition, for the $\tau\to\mu$ channel, the charge measurement status~\cite{internal_note:2009} of the daughter muon (negative or unknown) is also used. 

\paragraph*{Expected Events} \label{par:event_selection}
The expected number of $\nu_\tau$ events has been evaluated using the simulated CNGS flux~\cite{Ferrari:CNGS, CNGS:flux}, normalised to the number of observed $\nu_\mu$~CC interactions as explained in~\cite{OPERA:2tau}, assuming a maximal mixing ${\sin^{2}2\theta_{23} = 1}$, $\Delta m^{2}_{23} = 2.50\cdot 10^{-3}$~eV$^{2}$~\cite{PDGReview:2016} and the $\nu_\tau$ cross-section as in the default implementation provided by GENIE~v2.6~\cite{genie_paper,genie_tools}. 
The expected number of signal and background events are reported in Table~\ref{tab:NExp_MinBias}, together with the number of observed $\nu_\tau$ candidates for each channel.
\tabnexpectedevents
The background from $\pi$ and $K$ decays remain negligible.\\
The total systematic uncertainty on the expected signal, largely dominated from the limited knowledge of the $\nu_\tau$ cross-section and the detection efficiency, is conservatively set to $20\%$. 
Since signal expectation is calculated by using data-driven estimates of location efficiencies, this value is at first order insensitive to systematic effects on efficiencies up to the primary vertex location level.

Using the measured sample of $\nu_{\mu}$~CC interactions with charm production, the uncertainty on the charm background has been estimated to be about $20\%$~\cite{Agafonova:2014khd}. Hadron re-interaction background has an estimated uncertainty of $30\%$ from measurements of test-beam pion interactions in the OPERA bricks~\cite{Ishida:2014qga}. The systematic uncertainty on LAS has been obtained by a comparison between two different estimates, one based on a data-tuned GEANT4 Monte Carlo simulation~\cite{Longhin:2015dsa} and the other one on a direct extrapolation of data in the literature~\cite{Masek:1961zz} and it is set at $50\%$.\\
The total expected signal is ${N^{\textnormal{expS}}=(6.8\pm 0.75)}$ events, whereas the total background expectation is ${N^{\textnormal{expB}}=(2.0\pm0.4)}$ events.

\begin{figure}[h]
\centering
\includegraphics[scale=0.45]{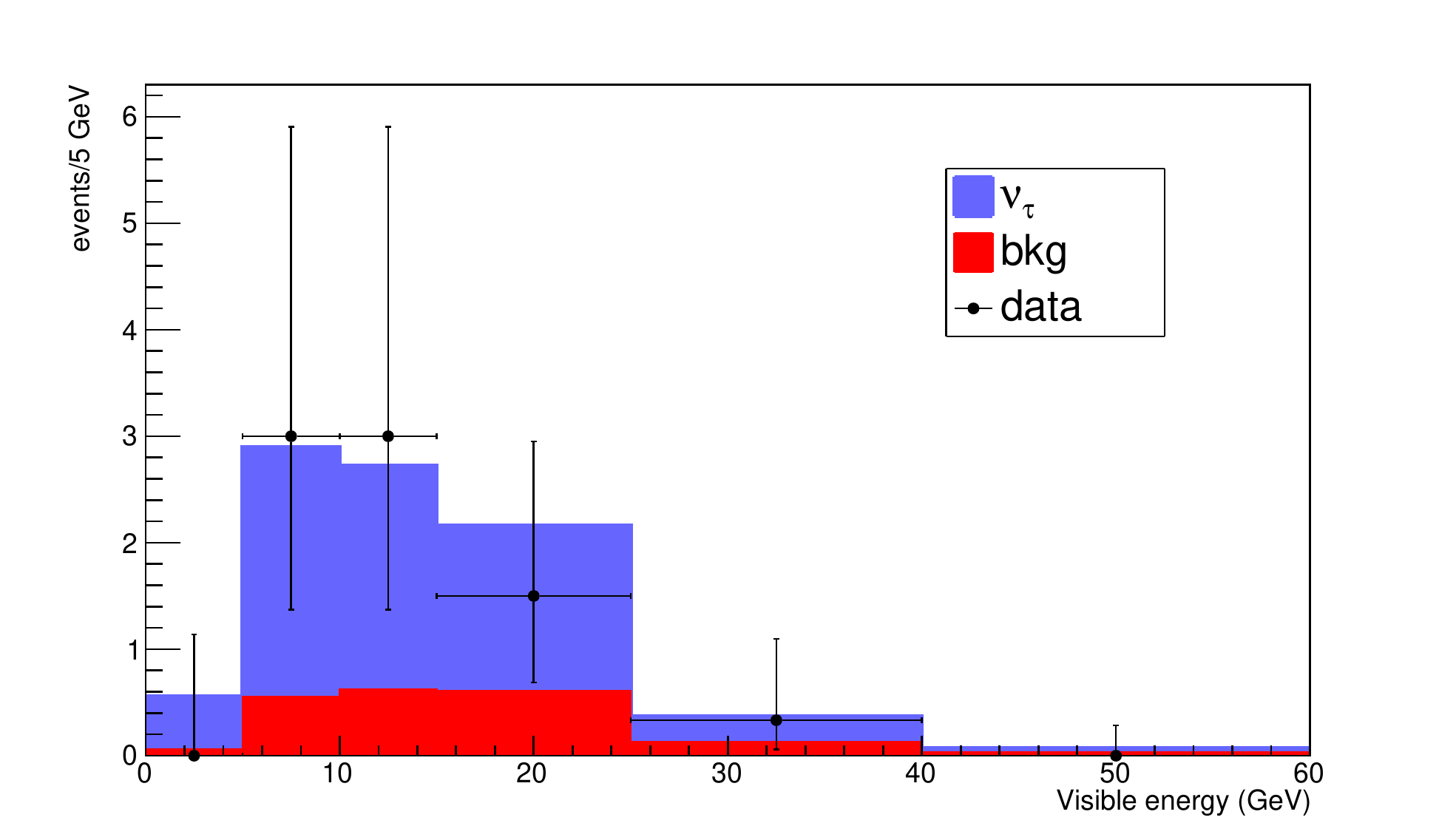}
\caption{Stacked plot of visible energy: data are compared with the expectation. Monte Carlo simulation is normalised to the expected number of events reported in Table~\ref{tab:NExp_MinBias}.}\label{fig:psum}
\end{figure}

\begin{figure}[h]
\centering
\includegraphics[width=88mm]{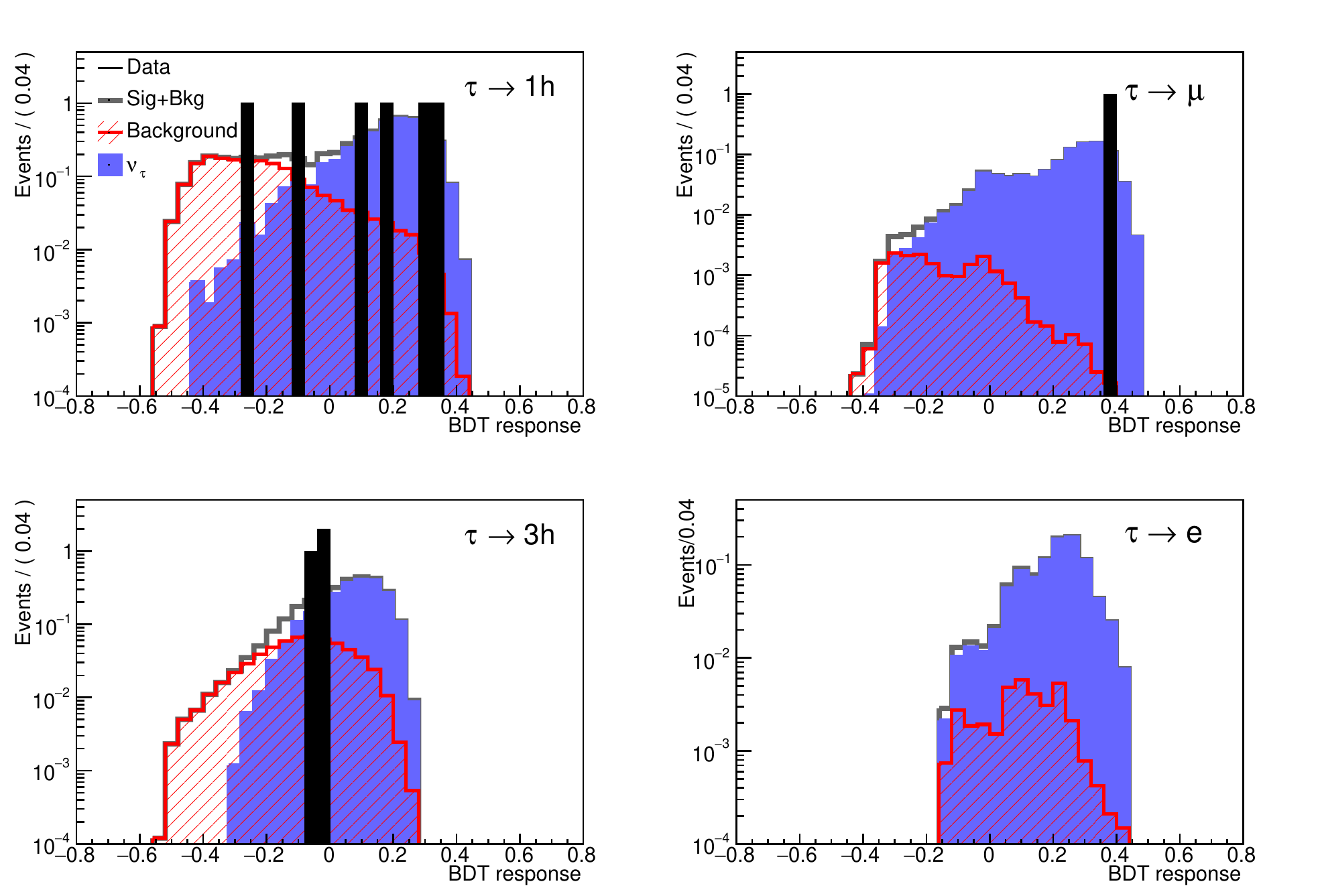}
\caption{BDT response for each channel.}\label{fig:BDT}
 \end{figure}

\paragraph*{Observed events}
Ten events ($N^{\textnormal{obs}}$) passed all the topological and kinematical cuts. The distribution of their visible energy, i.e.~the scalar sum of the momenta of charged particles and $\gamma$s, is shown in Fig.~\ref{fig:psum}, compared to Monte Carlo simulation.
Among the ten selected $\nu_\tau$ candidates, five $\nu_\tau$ were already described in~\cite{OPERA:1tau,OPERA:2tau,OPERA:3tau,OPERA:4tau,OPERA:5tau}.  
The other five $\nu_\tau$ candidates are all events without muon in the final state: three of them show a 1-prong decay and two a 3-prong decay. 
Their kinematical variables are summarised in Table~\ref{tab:MarginalEv}, where the BDT response for each event is also reported. 
The resulting BDT output distributions are shown in Fig.~\ref{fig:BDT}. 

\tabkinematicalvars

\section{Results\label{sec:results}}

The statistical analysis of the data employs a maximum-likelihood fit jointly across the four channels. 
For each channel, the likelihood is constructed as the product of a probability density function combining the BDT responses of signal and background, a Poissonian term and a Gaussian term for the systematics of the expected yield:
\begin{equation}\label{eq:LR}
\begin{split}
\mathcal{L}(\mu,\beta_{c}) & = \prod_{c=1}^{4}\left(\textnormal{Pois}(n_{c}|\mu s_{c} + \beta_{c}) \prod_{{i}=1}^{n_{c}}f_c(x_{ci})\right) \cdot
\\
& \qquad \cdot \prod_{c=1}^{4}
\textnormal{Gauss}(b_{c}|\beta_{c},  \sigma_{b_{c}}),
\end{split}
\end{equation}
where
$$
 f_c(x_{ci}) = \frac{\mu s_c}{\mu s_{c} + \beta_{c}} \mathrm{PDF}^{\mathrm{sig}}_{c} +
  \frac{\beta_{c}}{\mu s_{c} + \beta_{c}} \mathrm{PDF}^{\mathrm{bkg}}_{c},
$$
and $c$ runs over the 4 channels, $i$ over the $n_{c}$ observed events in the $c^{th}$ channel, $s_c$ is the expected signal, $b_c$ and $\sigma_{b_{c}}$ are the expected background in the c$^{th}$ channel and its uncertainty as reported in Table~\ref{tab:NExp_MinBias}, $\beta_c$ is a floating parameter which represents the true background, $x_{ci}$ is the BDT response, and $\mathrm{PDF}^{\mathrm{bkg}}_c$ ($\mathrm{PDF}^{\mathrm{sig}}_c$) the distribution of $x_{ci}$ for the background (signal) component in the  $c^{th}$ channel. The parameter $\mu$ is the $\nu_\tau$ signal strength, i.e.~a scale factor on the number of events expected by the model of neutrino interactions: $\mu=0$ corresponds to the background-only hypothesis and $\mu=1$ corresponds to the oscillated $\nu_\tau$ signal, on top of the background, reported in Table~\ref{tab:NExp_MinBias}. The effect of uncertainties on the expected number of signal events (estimated $\sim$20$\%$ for each channel) has been proved to be negligible for all the following results.

\paragraph*{Significance of $\nu_\tau$ appearance}
The significance of $\nu_\tau$ appearance is expressed in terms of a hypothesis test where the background-only ($\mu=0$) is the null hypothesis and the signal-plus-background ($\mu\neq0$) is the alternative one. 
In order to test which values of the signal strength $\mu$ are consistent with data, the profile likelihood ratio $\lambda(\mu) = \mathcal{L}(\mu, \hat{\hat{\beta}}_{c}(\mu))/\mathcal{L}(\hat{\mu}, \hat{\beta}_{c})$ is used~\cite{PDG:2016}, where $\mathcal{L}(\hat{\mu}, \hat{\beta}_{c})$ is the value of the likelihood at its maximum and $\hat{\hat{\beta}}_{c}(\mu)$ indicates the profiled values of the nuisance parameter $\beta_{c}$, maximizing $\mathcal{L}$ for the given $\mu$. 
The results presented in this letter are obtained using the asymptotic approximation~\cite{Cowan:2010js}, as implemented in the RooStats package~\cite{RooStats}.
 
The null hypothesis is excluded with 6.1~$\sigma$ significance, corresponding to a background fluctuation probability of $4.8 \cdot 10^{-10}$. 
The best-fit signal strength is ${\mu = 1.1^{+0.5}_{-0.4}}$, which is consistent with the $\nu_\tau$ appearance expected from neutrino oscillation. 

\paragraph*{First measurement of $\Delta m^2_{23}$ in appearance mode}

The $\nu_\tau$ signal strength $\mu$ is proportional to the oscillation probability and the $\nu_\tau$ cross-section. Assuming maximal mixing and $\nu_\tau$~CC interaction cross section as in previous section, the following interval of $\Delta m^2_{23}$ can be derived using the Feldman-Cousins method~\cite{Feldman:1997qc}: 
\begin{equation}
\Delta m^2_{23} = (2.7^{+0.7}_{-0.6})\cdot 10^{-3}\ \textnormal{eV}^{2}\; \textnormal{at}\ 68\%\ \textnormal{C.L.}
\end{equation}
This is the first result obtained in appearance mode and it is consistent with the disappearance results from different experiments, including the world average~\cite{PDG:2016}.

\paragraph*{Measurement of the $\nu_\tau$~CC cross-section}

Alternatively to the above measurement of $\Delta m^2_{23}$, one may fix $\Delta m^2_{23}$ at the world average value $(2.50\cdot 10^{-3}$~eV$^{2}$) and maximal mixing $\sin^2 2\theta_{23}=1$ and estimate the $\nu_\tau$~CC cross-section on the lead target, made of $^{204}$Pb (1.4$\%$), $^{206}$Pb (24.1$\%$), $^{207}$Pb (22.1$\%$) and $^{208}$Pb (52.4$\%$)~\cite{Anokhina:2008yy}.\\
The total flux integrated cross-section is defined as~\cite{Katori:2016yel}:
\begin{equation}
\langle  \sigma \rangle = \frac{\int \Phi_{\nu_\mu}(E) \mathcal{P}_{\nu_\mu\to\nu_\tau}(E)  \sigma_{\nu_\tau}(E) dE }{\int \Phi_{\nu_\mu}(E) \mathcal{P}_{\nu_\mu\to\nu_\tau}(E) dE},
\end{equation}
where $\Phi_{\nu_\mu}(E)$ is the CNGS flux~\cite{CNGS:flux}, $\mathcal{P_{\nu_\mu\to\nu_\tau}}$ the oscillation probability, $ \sigma_{\nu_\tau}(E)$ is the $\nu_\tau$ cross-section and E is the neutrino energy.

An estimate of $\sigma_{\nu_\tau}$ can be extracted 
from the observed data by using the following equation:
\begin{equation}
\langle  \sigma \rangle_{meas} = \frac{(N^{obs} -N^{expB})/{(\epsilon N_T )}}{\int \Phi_{\nu_\mu}(E) \mathcal{P}_{\nu_\mu\to\nu_\tau}(E) dE},
\end{equation}
where $N_T$ is the number of lead nuclei in the fiducial volume of the target and $\epsilon$ is the overall efficiency of $\tau$ event reconstruction, averaged over the expected distribution of $\nu_\tau$ flux~\cite{nutauflux:operapubnote186}. The result is:
\begin{equation}
\langle  \sigma \rangle_{meas} =  (5.1^{+2.4}_{-2.0}) \cdot 10^{-36} \textnormal{cm}^2,
\end{equation}
with the error dominated by statistical uncertainty. 
\begin{figure}
\centering
\includegraphics[scale=0.45]{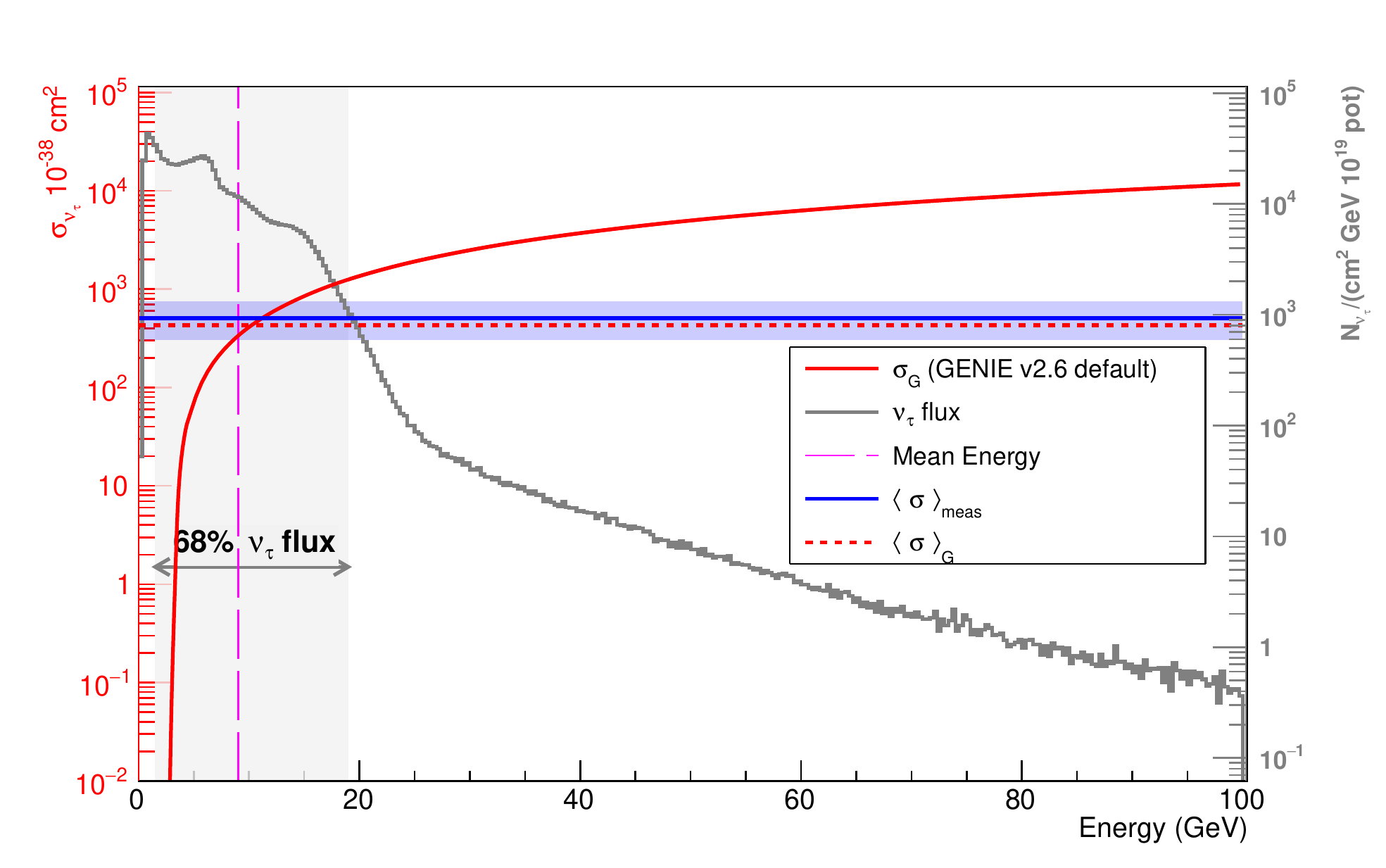}
\caption{Flux-averaged measurement of the CC~$\sigma_{\nu_\tau}$ on a lead target. 
The horizontal bar includes central 68$\%$ of the flux.}\label{fig:nutau_crosssection_plot_SKlike}
\end{figure}
This result has to be compared with the expected value, as provided by the default configuration of GENIE v2.6~\cite{genie_paper,genie_tools}: ${\langle \sigma \rangle_{G}=(4.29\pm0.04)\cdot 10^{-36} \textnormal{cm}^2}$. The error associated with $\langle \sigma \rangle_{G}$ is the propagation of the flux uncertainty due to the oscillation parameter errors. Therefore, the result can be expressed as: ${\langle \sigma \rangle_{meas} = (1.2^{+0.6}_{-0.5})\ \langle \sigma \rangle_{G}}$. 

This is the first measurement of the $\nu_\tau$~CC cross-section with a negligible contamination of $\bar{\nu}_\tau$. No deviation from GENIE expectations is observed.

\paragraph*{$\nu_\tau$ lepton number}

The lepton number of $\nu_\tau$ has never been observed. In the muonic channel, the OPERA experiment can distinguish neutrinos from anti-neutrinos looking at the charge of the muon produced in $\tau$ decays, which, was measured to be negative at 5.6~$\sigma$ level for the $\tau\to\mu$ candidate~\cite{Longhin:operapubnote161,OPERA:3tau}.\\
The hypothesis that a ${\tau^{-}\to\mu^{-}}$ has been observed is tested by specifying Eq.~\ref{eq:LR} to the  $\tau\to\mu$  channel. A dedicated BDT analysis has been performed for this channel: on top of the charm and LAS background, we have considered the additional contribution from the 2$\%$ contamination in interactions of $\bar\nu_\tau$ resulting from $\bar\nu_\mu$ oscillation. Interactions of $\bar\nu_\tau$ are background in the muonic channel if the $\mu^+$ charge is either misidentified or not measured: this gives a background yield of ${0.0024\pm0.0005}$ events.\\
The result gives a significance of 3.7~$\sigma$, which, assuming the lepton number conservation in the neutrino interaction, can be considered as the first direct observation of the $\nu_\tau$ lepton number.

\section{Conclusions}
This letter reports OPERA's final results on $\nu_\mu\rightarrow\nu_\tau$ oscillations in appearance mode, obtained with the complete data sample, corresponding to 5603 $\nu$ interactions fully reconstructed.

Given the validation of the Monte Carlo simulation of $\nu_\tau$ events, based on different control data samples, a new analysis strategy was developed, fully exploiting the features expected for $\nu_\tau$ events. A multivariate approach for the identification of $\nu_\tau$ events was applied to candidate events selected by means of moderately tight topological and kinematical cuts.

Ten~$\nu_\tau$ candidates were observed, with~${2.0\pm0.4}$ expected background events. 
The discovery of $\nu_\mu\to\nu_\tau$ oscillations in appearance mode is confirmed with an improved significance of 6.1~$\sigma$.

Assuming ${\sin^2 2\theta_{23}=1}$, the first measurement of $\Delta m^2_{23}$ in appearance mode gives ${(2.7^{+0.7}_{-0.6})\cdot 10^{-3}\textnormal{eV}^{2}}$, while the $\nu_\tau$~CC cross-section on the lead OPERA target is measured to be ${(5.1^{+2.4}_{-2.0}) \cdot 10^{-36} \textnormal{cm}^2}$, when ${\Delta m^2_{23}=2.50\cdot 10^{-3}\ \textnormal{eV}^{2}}$. 

Furthermore, a dedicated BDT analysis in the $\tau\to\mu$ channel allows claiming for the first direct observation of the $\nu_\tau$ lepton number with a significance of 3.7~$\sigma$.

\section*{Acknowledgements}
\footnotesize{We warmly thank  CERN for the successful operation of the CNGS facility and INFN for the continuous support given by hosting the experiment in its LNGS laboratory. Funding is gratefully acknowledged from  national agencies and Institutions supporting us, namely: Fonds de la Recherche Scientifique-FNRS and Institut Interuniversitaire des Sciences Nucleaires for Belgium; MoSES for Croatia; CNRS and IN2P3 for France; BMBF for Germany; INFN for Italy; JSPS, MEXT, the QFPU-Global COE program of Nagoya University, and Promotion and Mutual Aid Corporation for Private Schools of Japan for Japan; SNF, the University of Bern and ETH Zurich for Switzerland; the Russian Foundation for Basic Research (Grant No. 12-02-12142 ofim), the Programs of the Presidium of the Russian Academy of Sciences (Neutrino Physics and Experimental and Theoretical Researches of Fundamental Interactions), and the Ministry of Education and Science of the Russian Federation for Russia, the Basic Science Research Program through the National Research Foundation of Korea (NRF) funded by the Ministry of Science and ICT (Grant No. NRF-2018R1A2B2007757) for Korea; and TUBITAK, the Scientific and Technological Research Council of Turkey for Turkey (Grant No. 108T324).}

\bibliography{xbib}

\end{document}